%% file: aa.tex
\documentclass{aa}  

\usepackage{graphicx}
\usepackage{txfonts}
\usepackage{multirow}
\usepackage{hyperref}
\usepackage{breakurl}
\newcommand{\chandra}{{\it Chandra}}
\newcommand{\swift}{{\it Swift}}
\newcommand{\xmm}{{\it XMM-Newton}}

\begin{document}

\title{Early Emergence of SSS emission in RS\,Oph}

   \author{J.-U. Ness\inst{\ref{esa}}\and
  J.P. Osborne\inst{\ref{leicester}}\and
  M. Orio\inst{\ref{wisconsin},\ref{padova}}\and
  S. Starrfield\inst{\ref{asu}}\and
  S. Mitrani\inst{\ref{telaviv}}\and
  E. Behar\inst{\ref{telaviv}}
    }

\institute{European Space Agency (ESA), European Space Astronomy Centre (ESAC), Camino Bajo del Castillo s/n, 28692 Villanueva de la Ca\~nada, Madrid, Spain; corresponding author: \email{jan.uwe.ness@esa.int}\label{esa}
\and
School of Physics \& Astronomy, University of Leicester, Leicester, LE1 7RH, UK\label{leicester}
        \and
Dept. of Astronomy, University of Wisconsin, 475 N, Charter Str., Madison, WI 53706, USA\label{wisconsin}
     \and
INAF-Osservatorio di Padova, Vicolo Osservatorio 5, 35122 Padova, Italy\label{padova}
      \and
    School of Earth and Space Exploration, Arizona State University, Tempe, AZ 85287-6004, USA\label{asu}
    \and
    Department of Physics, Technion, Haifa 32000, Israel\label{telaviv}
}
   \authorrunning{Ness et al.}
   \titlerunning{2021 outburst of RS\,Oph in X-rays}
   \date{Received \today; accepted }

 
  \abstract
{
Early in the 2006 outburst of the recurrent nova RS\,Oph, \swift\ detected the onset of the super-soft-source (SSS) phase. An \xmm\ observation obtained on day 26.1 revealed a highly unusual soft X-ray spectrum containing transient narrow emission features that could not be identified with known transitions and are atypical of SSS spectra. Here we analyse a new \xmm\ observation obtained at a comparable stage of the 2021 outburst.
}
{
By means of presenting a complete census of temporal and spectral properties, we aim to demonstrate the high degree of complexity, deriving requirements for future model approaches.
}
{
To minimise model-dependent bias, we use models only for demonstration purposes and base the interpretation primarily on observational evidence.
We use the 2006d26.1 EPIC/pn spectrum to derive a minimal model for the onset of SSS emission and compare it with the simultaneous RGS spectrum. We investigate the detailed properties of the 2021d37.1 RGS spectrum through comparisons with spectra from other epochs of the 2006 and 2021 RS\,Oph outbursts, spectra extracted from four variability intervals within the observation, and spectra of other novae with similar characteristics. These comparisons are used to identify plausible emission mechanisms and guide spectral interpretation. We also search for 35-s periodic signals in the 2021d37.1 RGS and simultaneous NICER data, motivated by their detection during the 2006d26.1 observation.
}
   {
A blackbody model reproduces the low-resolution 2006d26.1 EPIC/pn spectrum but fails to describe the simultaneous RGS data. The 2021d37.1 RGS spectrum closely resembles that from 2006d26.1 and consists of a slowly fading shock component and a highly variable soft component with several unidentified features. While the shock emission evolves on timescales of days, the soft component varies on hourly timescales and dominates the observed variability, with individual features showing distinct behaviour. The soft emission resembles that of the novae V1716\,Sco and V1494\,Aql, the latter having recently been interpreted with a charge-exchange model, but not that of the persistent SSS Cal\,87. No significant 35-s periodicity is detected.
}
{
Our results demonstrate that low spectral-resolution observations can fail to reveal the complex line-dominated emission seen by higher resolution instruments.
The rapidly variable soft component in RS\,Oph, together with Radiative Recombination Continua (RRC) features, Fe\,{\sc x} emission, and the previously detected 35-s period, suggests that a hidden photoionising source is present.
At the same time, the observed spectral complexity and variability indicate that quantitatively reproducing these spectra is likely to require modelling that accounts for geometry, density structure, dynamics, absorption, clumpiness, and other effects.
We therefore focus on empirical comparisons and conclude that, in the classification scheme of \citet{nessobsc}, the enigmatic spectra observed shortly before the emergence of a clear SSS spectrum form an intermediate class between SSe and SSa spectra, together with V1494 Aql and V1716 Sco. 
Future modelling efforts will be required to establish the relative importance of the physical processes responsible for this broader class of intermediate SSS spectra. Possible approaches include photoionisation models, which naturally arise from the presence of the luminous supersoft source, as well as charge-exchange models such as those recently proposed by \citet{mitrani_v1716sco} for V1716 Sco.
    }

   \maketitle

%

\section{Introduction}

\begin{table*}[ht]
\begin{flushleft}
\renewcommand{\arraystretch}{1.1}
\caption{\label{tab:obs}Journal of exposures taken with \xmm\ and \chandra\ (last rows) used in this work}
\begin{tabular}{lllllll}
\hline
ObsID & Instr. & Mode & Filter & Start time & Stop time & exposure time\\
\hline
\input{logtab.tex}
\end{tabular}
$^a$Observation start in days after $t_{\rm ref}$=2021-08-09.5417
$^b$Observation start in days after 2006-02-12.84
$^c$V1494\,Aql, day 304 after optical maximum
$^d$V1716\,Sco, day 127 after discovery
$^e$Cal\,87
\renewcommand{\arraystretch}{1}
\end{flushleft}
\end{table*}

Classical novae occur in accreting binary systems containing a white dwarf
and a mass-donor companion. Hydrogen-rich material accumulates on the
white-dwarf surface until thermonuclear runaway ignites near the base of
the accreted layer. The resulting energy release initially emerges at high
energies, while radiation pressure drives the expansion and ejection of the
envelope. The earliest fireball phase, during which soft X-rays can escape
directly from the photosphere, is expected to be short-lived and has only
been unambiguously observed once, by \emph{eROSITA}
\citep{ole}.\\

As the ejecta expand, their changing opacity governs the observed
evolution. The nova first reaches maximum optical brightness while the
atmosphere remains extended and optically thick. Continued expansion
causes the photosphere to recede to smaller radii and higher effective
temperatures, shifting the spectral energy distribution towards higher
energies. This evolution culminates in the super-soft-source (SSS) phase,
characterised by bright soft X-ray emission with effective temperatures of
a few $10^5$\,K. The SSS phase persists until nuclear burning ceases and
can last from weeks to months. Persistent SSS systems, such as Cal\,83,
are thought to maintain a balance between accretion and nuclear burning
\citep{heuvel}.\\

At some point after the eruption, accretion resumes and may eventually
trigger a new outburst. Novae with multiple \emph{observed} eruptions are
referred to as recurrent novae. RS\,Oph is both a recurrent and a
symbiotic nova, the latter characterise by its donor star being a red giant in a long-period orbit of
$453.6 \pm 0.4$\,days \citep{brandi08}. The dense stellar wind of the
giant provides an environment markedly different from that of novae with
main-sequence companions. Confirmed outbursts have occurred roughly every
10--20 years since 1898. The most recent eruption was reported by the
AAVSO on 2021 August 9, reaching an optical peak magnitude of 4.5 around
2021-08-09.5417 UT, which we adopt as reference time $t_{\rm ref}$.\\

The dense red-giant wind interacts with the expanding nova ejecta,
converting part of their kinetic energy into thermal energy and producing
shock-powered X-ray emission. This component evolved similarly during the
2006 and 2021 eruptions \citep{bode06,rsophshock,orio2022,orio23}. The
interaction also leaves signatures in high-resolution SSS spectra,
including reduced absorption-line blue shifts and increasing ionisation
with time, indicating that the shocks heat the ejecta in addition to
producing hard X-rays \citep{ness2023}.\\

Both the 2006 and 2021 outbursts were extensively observed with the same
fleet of X-ray instruments: Long-term monitoring with low spectral resolution
instruments was carried out with
\swift\ \citep{osborne11,page2022,orio23}, while deeper observations
at high spectral resolution were obtained at lower cadence with \chandra\ and \xmm\
\citep{nessrsoph,nelson07,orio2022,ness2023}. These observations revealed
that the observed SSS brightness can depend sensitively on obscuration.
For example, the two recent eruptions of RS\,Oph
showed markedly different soft X-ray luminosities despite similar
behaviour at harder X-ray, optical, and radio wavelengths
\citep{page2022}. High-resolution spectroscopy suggests substantially
stronger absorption of the soft X-rays in 2021 than in 2006
\citep{ness2023}. Since photoelectric absorption is particularly effective
at soft X-ray energies, relatively small changes in absorbing column
density can dramatically alter the observed flux. Consequently, the onset
of observable SSS emission from Earth need not coincide with the first appearance
of SSS emission, and a local SSS source may already be present while remaining hidden
when viewed from Earth.\\

Several observations indicate that ionising SSS emission was already
present before its direct detection in RS\,Oph. First signs include
transient soft excesses observed between days $\sim20-26$ in the
\swift\ and NICER monitoring data of both the 2006 and 2021 eruptions
\citep{page2022,orio23}. An \xmm\ observation obtained on day~26.1
during the 2006 eruption resolved this soft emission into a complex
high-resolution spectrum that does not appear like other typical
SSS spectra. This peculiar spectrum was dominated by structured features
between the strong N\,{\sc vii} and N\,{\sc vi} lines at 24.78 and 28.78\,\AA,
respectively \citep{nessrsoph,nelson07}. The transient nature of these
features was demonstrated by \cite{basi} and \cite{nessobsc}.\\

Independent observational evidence for hidden SSS emission is provided by optical coronal lines. While an SSS source may be hidden from view owing to a high degree of photoelectric absorption, highly ionised forbidden lines of Fe can emerge through material that is opaque to soft X-rays and are observable from the ground at optical wavelengths. Ionisation stages up to Fe,{\sc ix} can be produced by EUV/very-soft X-ray radiation\footnote{The ionisation energy from Fe,{\sc viii} to Fe,{\sc ix} is 151.06,eV ($\lambda<82.1$,\AA)}, whereas production of Fe,{\sc x} requires photons with energies $>233.6$,eV ($\lambda<53$\,\AA), i.e. in the soft X-ray regime. During both the 2006 and 2021 eruptions, the [Fe,{\sc x}] emission evolved very similarly despite the large difference in directly observed soft X-ray fluxes. Furthermore, the sharp rise in [Fe,{\sc x}] preceded the appearance of bright SSS emission in the \swift\ light curves by several days (see figure 14 in \citealt{page2022}). Since the presence of these coronal lines requires a substantial flux of ionising photons, their evolution suggests that SSS emission was already present while still obscured along our line of sight.\\

Additional support comes from the transient short-period oscillation
discovered in the 2006 eruption. A quasi-periodicity near 35\,s was
detected only during the SSS phase and only in the soft X-ray component,
not in the harder shock emission
\citep{atel770,osborne11,nessrsoph,nessqpo}. Its brief appearance during
the flare in the day-26.1 \xmm\ observation therefore associates the
peculiar spectrum at that epoch with the emergence of SSS emission, even
though the high-resolution spectrum did not resemble the absorption-line
dominated atmosphere spectra typically observed during fully developed
SSS phases.\\

High-resolution SSS spectra display a remarkable diversity of
morphologies. While many are dominated by continuum emission and
absorption lines, others exhibit strong emission-line spectra
superimposed on a weak continuum. To describe this diversity,
\cite{nessobsc} introduced the classes SSa and SSe. SSa spectra are
dominated by continuum emission and atmospheric absorption lines, whereas
SSe spectra are characterised by prominent emission lines with only a
weak continuum contribution. Examples of SSe behaviour include the
recurrent nova U\,Sco and the persistent SSS Cal\,87, both of which are
high-inclination systems. This led \cite{nessobsc} to propose that
geometrical obscuration of the central continuum source enhances the
relative visibility of line-emitting material. In this picture, SSa and
SSe spectra may represent different levels of obscuration and
reprocessing rather than strictly distinct physical classes, allowing
intermediate cases to occur.\\

The physical origin of the emission features in such structured soft
X-ray spectra remains debated. A central SSS source naturally provides an
intense field of ionising photons, making photoionisation an obvious
mechanism to consider. Indeed, the [Fe\,{\sc x}] evolution in RS\,Oph
constitutes compelling evidence that photoionising radiation is present
even when the central source is not directly observable. However,
photoionisation may not provide a unique explanation for all of the
spectral features observed during the emergence of the SSS phase.\\

Recent work has demonstrated that charge-exchange processes can produce
many of the characteristic features seen in high-resolution nova spectra.
Using \xmm\ and \chandra\ observations of YZ\,Ret, \cite{mitrani2024}
identified narrow radiative recombination continua and enhanced
high-$n$ transitions that they interpreted as evidence for charge
exchange between hot shocked plasma and cold gas. A similar framework was
applied by \cite{mitrani_v1716sco} to Nova Sco~2023 (V1716\,Sco), where
the observed emission-line spectrum was successfully described by a model
involving interactions between highly ionised plasma and cool material.
These studies demonstrate that radiative recombination continua and
complex emission-line spectra do not uniquely imply a photoionised
plasma and motivate consideration of alternative physical mechanisms in
the interpretation of structured nova spectra.\\

The emergence of SSS emission in RS\,Oph does therefore not necessarily
signal the simple clearing of the obscuration of the nuclear burning source
but may instead reflect changing observability of an already active
central source. The early soft X-ray spectra obtained during this
transition exhibit complex structure that is not straightforwardly
explained within the framework of a classical absorption-line dominated
SSS atmosphere spectrum. They must therefore be interpreted
within a framework that potentially includes direct but obscured SSS
emission, reprocessing by surrounding material, ejecta-wind interactions,
and possible charge-exchange processes at interfaces between hot and cold
plasma. In this work we investigate these transitional
high-resolution X-ray spectra in the context of the overall evolution of
the 2006 and 2021 eruptions. We examine their relation to the emergence
of observable SSS emission, compare their properties with the SSa and SSe
spectral phenomenology, and discuss possible physical mechanisms
responsible for the observed emission features. The observations and data
reduction are described in Sect.~\ref{sect:obs}, the results are
presented in Sect.~\ref{sect:results}, perspectives for future modelling
are described in Sect.~\ref{sect:perspectives}, and our conclusions are summarised
in Sect.~\ref{sect:concl}.

\section{Observations}
\label{sect:obs}

The details of all exposures taken on day 2021d37.1 are listed in Table~\ref{tab:obs} followed by the exposure details of RGS exposures for two archival \xmm\ observations taken days 2006d26.1 and 2006d54.0 and the \chandra\ observation taken on day 2006d39.7.\\

\begin{figure*}[!ht]
\resizebox{\hsize}{!}{\includegraphics{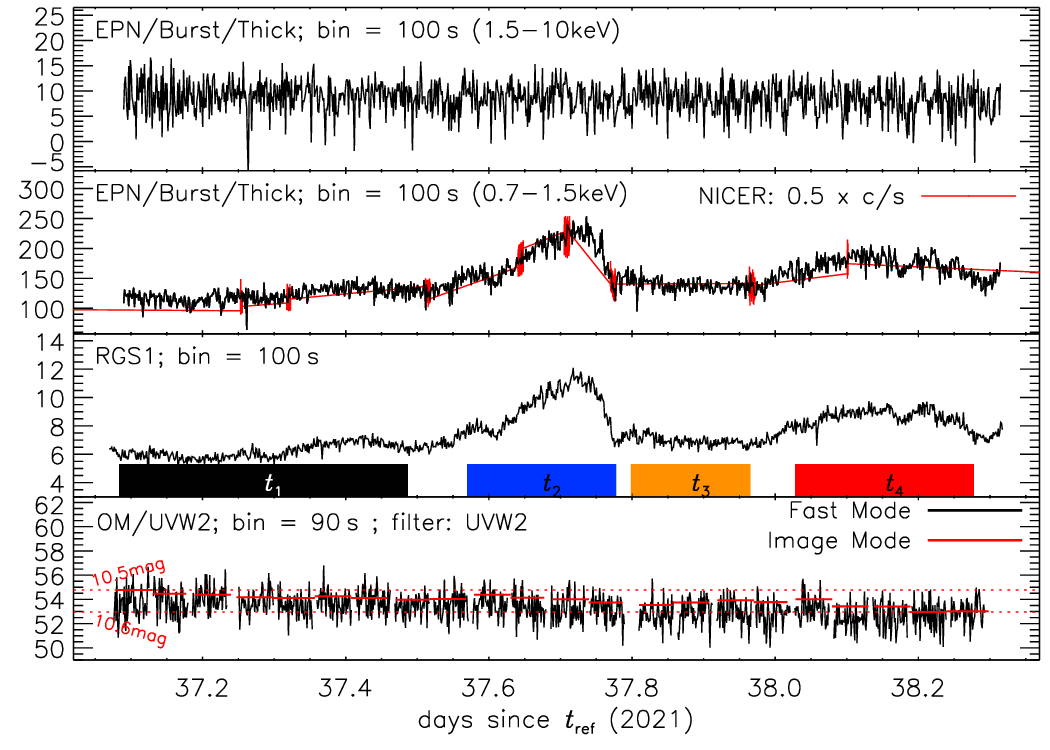}}
	\caption[multilc]{\label{fig:xmmlc}\xmm\ X-ray and UV light curves (units counts per second, cps, in all panels) extracted from the pn, RGS1, and Optical Monitor (OM) instruments taken on day 2021d37.1. The photon arrival times were converted to days after $t_{\rm ref}$=2021-08-09.5417. The pn light curve is broken into a hard (top) and a soft (2nd panel) band, split at 1.5\,keV probing shock and SSS emission, respectively; since the pn was operated in burst mode and no MOS light curves are available (see Table~\ref{tab:obs}), we only have one soft EPIC light curve covering the 0.7-1.5\,keV range. We also include NICER count rates in the 2nd panel, scaled down by a factor 2 \citep{orio23}. The RGS light curve (3rd panel) extends further to soft energies (0.3-2\,keV), and the shaded coloured area in the 3rd panel marks light curve segments discussed separately in \S\ref{sect:results:specvar}. In the bottom panel, the OM UVW2 filter ($\sim 2070-2170$\,\AA\ range) light curves are shown from the Image mode (red horizontal bars for each exposure) and Fast mode (black) data. Corresponding magnitude values are included in the left that correspond to the lowest and highest Image mode count rates (marked with the dotted horizontal red lines).}
\end{figure*}

For the observation to be taken on day 2021d37.05, we expected an extremely bright soft source (like on day 2006d39.7), and thus selected the RGS Small Window mode where the active chip size is reduced in order to increase the readout speed by a factor of 4 whereas both MOS instruments are deactivated to use their telemetry allocation to downlink all RGS data without disruptions. As a further precaution to prevent pile up and telemetry saturation, the pn was operated in the burst mode. For measurements of SSS emission, the burst mode is not ideal as it is limited to emission above 0.7\,keV. Therefore, any results obtained from the pn data are representative only of the hard tail of the SSS spectrum and the shock emission.\\

The uncalibrated observation data (Observation Data File=ODF) were downloaded from the ESA \xmm\ archive and processed with Science Analysis Software (SAS) version 19.0.0 and the latest CCF version. For the Optical Monitor (OM) data, the two standard chains {\tt omichain} and {\tt omfchain} were run to produce image and fast-mode data, respectively.\\

A standard {\tt rgsproc} run did not produce any events, and the \xmm\ Helpdesk reported an anomaly called "G7005 OOL during all observation" for the RGS1. Fully calibrated RGS spectra could still be produced with a workaround provided by the \xmm\ SOC that is described in the appendix \S\ref{appendix:g7005OOL}.\\

For qualitative illustration purposes we use the output of the SAS tool {\tt rgsfluxer} which is a product merging the RGS1 and RGS2 spectra into a single spectrum in flux units.
As described in \S\ref{appendix:g7005OOL}, {\tt rgsfluxer} had to be run with {\tt bg=''}, thus without background subtraction as the RGS Small Window mode contains no chip regions with pure background emission. Given the brightness of the source, we consider the background negligible.\\

Another issue described in appendix \S\ref{appendix:g7005OOL} is a GTI\&EXPOSU Inconsistency error having prevented the production of an RGS light curve with {\tt rgslccorr}. This ODF contains a different number of rows in the GTI and EXPOSU extensions. There is currently no solution and the \xmm\ SOC is working on a fix to be implemented in a future SAS release. As a workaround, we produced separate RGS1 and RGS2 light curves with {\tt evselect} and an optimised extraction region.\\

Any results obtained from one instrument would ideally be confirmed with data from at least one of the other instruments, and \xmm\ is ideal with so many instruments operating in parallel. Unfortunately, as described above, we have no MOS data and the pn data, limited to energies above 0.7\,keV, is not representative of the SSS emission. We can thus only compare the uncalibrated RGS1 and RGS2 light curves as independent measurements.\\

\section{Results}
\label{sect:results}

This section presents the observational results independently of a specific physical interpretation. The findings are summarised at the beginning of \S\ref{sect:perspectives}, where they provide the basis for evaluating possible interpretations.

\subsection{Variability Studies}
\label{sect:results:var}

In Fig.~\ref{fig:xmmlc}, the X-ray and UV light curves from day 2021d37.1 are shown,
extracted from the pn, NICER \citep{orio23}, RGS1, and OM exposures. The hard X-ray light curve whose
emission originates in the shocks is flat, demonstrating that all variability
originates in the SSS component. On day $\sim 37.6$, a $\sim 50-60$\%
increase in count rate can be seen in the soft pn and the RGS1 lightcurves.
Also towards the end of the observation, around day $\sim 38$, another, slower
increase can be seen.\\

In the soft X-ray light curves, we searched for the 35-s quasi-periodic
oscillations detected during the 2006 outburst. This period has again
been found with \swift\ during the 2021 outburst \citep{page2022,ness2023,orio23} indicating
that it must be related to system parameters such as mass of the white dwarf, that do
not change with each outburst. Since it was transient
in 2006, we studied the time evolution following the methods described
in \cite{nessqpo} thus splitting the pn and RGS light curves into adjacent
overlapping subintervals for which we calculated Lomb-Scargle power spectra
using the method of \cite{hornebal}. We did not detect reliable periodic variations
at any time during the 2021d37.1 observation. The RGS1 data yield the highest quality
light curve with the shorter readout time (compared to RGS2) and covering the
entire energy range (compared to pn): the RGS2 has twice the readout time,
and the pn only covers emission above 0.7\,keV.
We extracted light curves with different time bin sizes, and while detecting
periodic signals in some of these light curves, we find different results
for different time bin sizes, and for this reason we conclude that no
detectable period is present in the dataset taken on day 2021d37.1.
We also inspected the simultaneous NICER data that are shown in the second panel
in Fig.~\ref{fig:xmmlc} where it can be seen that NICER measured the same variability
trends. Unfortunately, the NICER light curve segments were too short for significant
detections of the 35s period. While we found elevated power values around 35s
in the power spectra at times around the peak in the light curve in both
\xmm\ and NICER light curves, these are not statistically significant and can be
coincidence noise.
Therefore, while on day 2006d26.1, the 35s was detected
at low significance during the brightness increase, we cannot confirm the same behaviour
for the day 2021d37.1.\\

\begin{figure}[!ht]
\resizebox{\hsize}{!}{\includegraphics{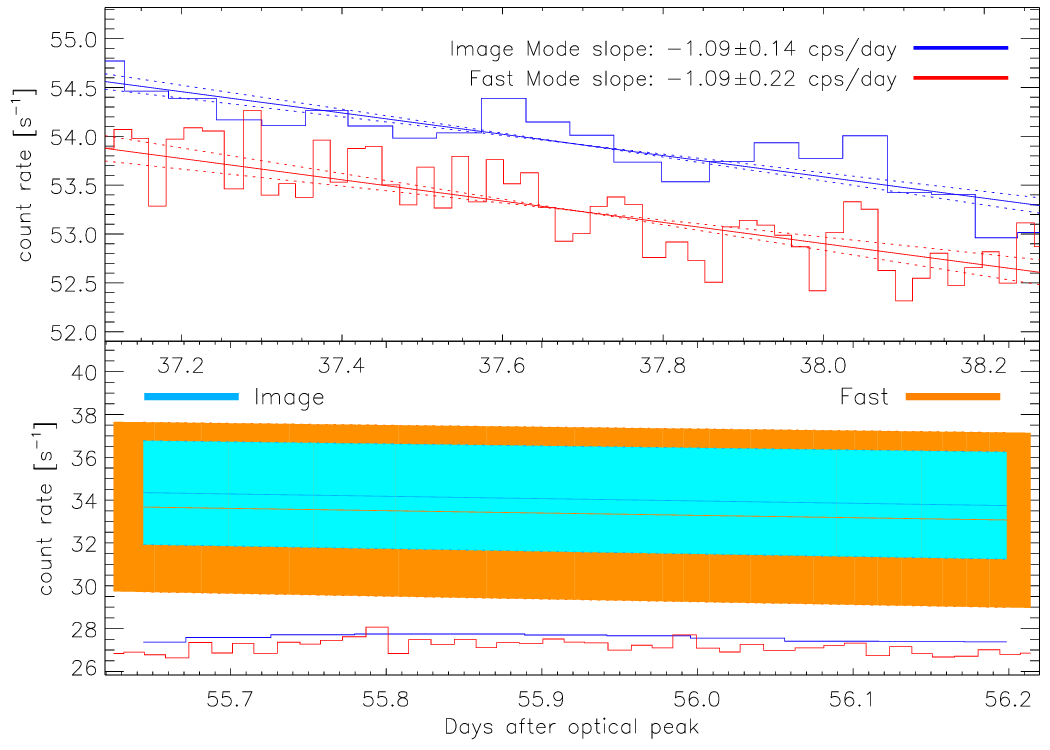}}
\caption{\label{fig:omfit}\xmm\ OM Image Mode (blue) and Fast Mode (red/orange)
light curves on day 2021d37.1 (top) and 2021d56 (bottom) with linear
fits to the 2021d37.1 light curves yielding a slow decline of $\sim 1$ count per second (cps) per day. The bottom panel shows the extrapolation of these fits with respective shaded areas (marking the 1-$\sigma$ uncertainty ranges) and the observed light curves being significantly lower than the extrapolations.
	}
\end{figure}

As can be seen from the bottom panel in Fig.~\ref{fig:xmmlc}, the pronounced
intensity variations seen in X-ray are not present in the OM/UVW2 band which
only varies within the narrow magnitude range 10.6-10.5. The red
dotted horizontal lines mark this range between days
37.07 to 38.3, and a slow systematic decline may be seen with
the first image mode magnitude being the highest while the last one yields
the lowest value. This systematic decline can better be seen in
Fig.~\ref{fig:omfit} where linear fits to the day 37.07-38.3 Image- and Fast-Mode
light curves are shown in the top, while the respective extrapolated count rates
expected to be seen in the day 55.62-56.22 range are shown in the bottom. The difference
is not large, and over a time interval of $\sim 20$ days the decline may have
been slightly faster or not quite linear. We cannot determine the decline pattern
from the Swift UVOT because the source was too bright for the UVW2 filter
($<10.95$\,mag) around the times of days 2021d37.1 and 2021d56.\\

\begin{figure}[!ht]
\resizebox{\hsize}{!}{\includegraphics{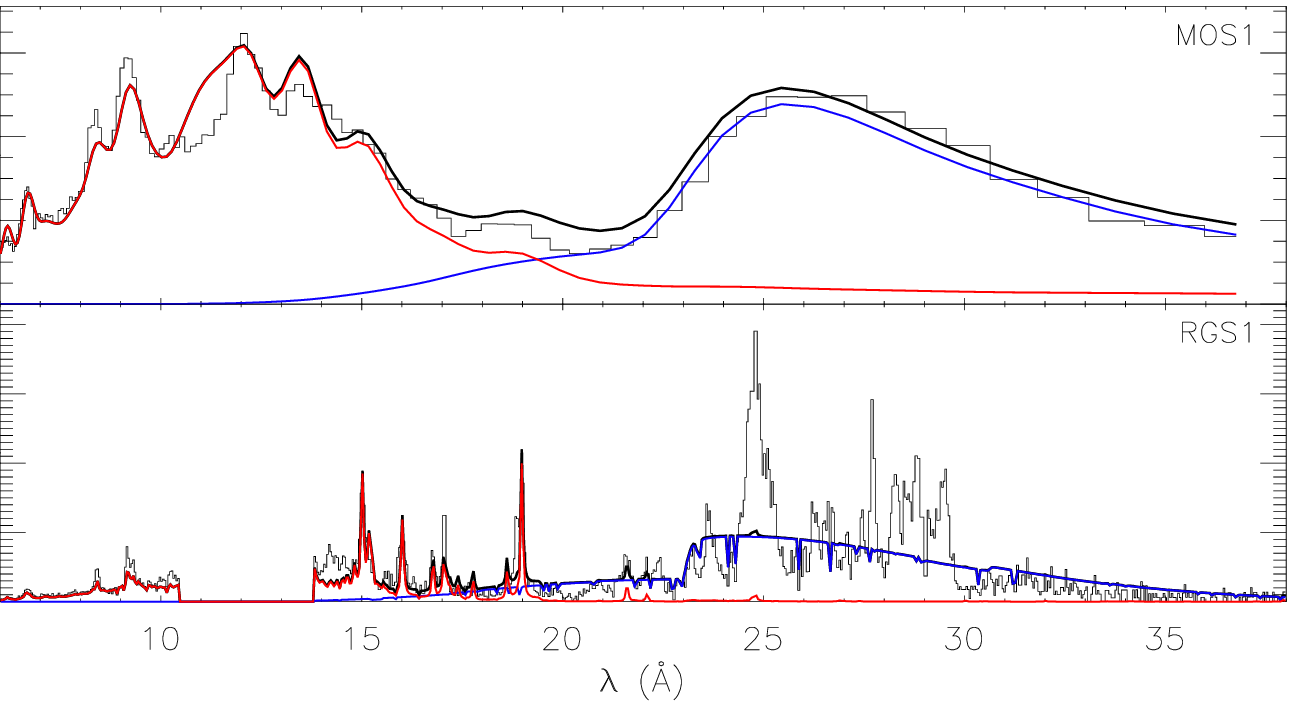}}
	\caption{\label{fig:epic_rgs}Comparison of EPIC/MOS1 (top) and RGS1 (bottom) spectra (in wavelength units) taken during the flare-like brightness increase on day 2006d26.1 having led to the soft component between 23\,\AA\ and 35\,\AA. The model (smooth black line) is the sum of a collisional ({\tt apec}, red line) and a blackbody (blue line) component. The blackbody component represents a good fit to the MOS1 spectrum while the RGS1 spectrum demonstrates that it is nothing like a blackbody: The model in the bottom panel is the same model as in the top panel but folded through the RGS1 instrumental response.}
\end{figure}

\begin{figure*}[!ht]
\resizebox{\hsize}{!}{\includegraphics{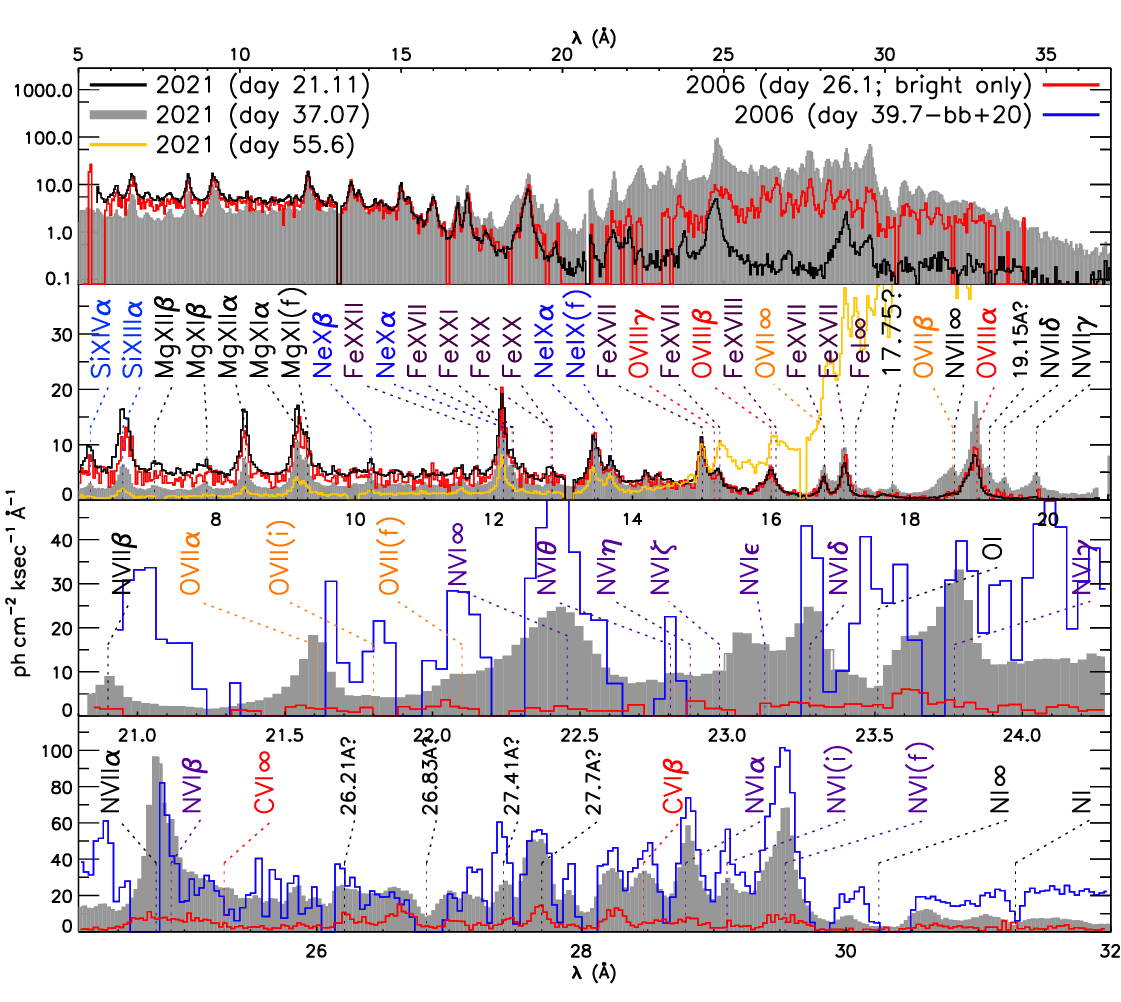}}
\caption{\label{fig:xmmspec37}\xmm\ merged and fluxed RGS1+RGS2 spectra in photon flux units per
10$^3$\,s (=ksec), ph\,cm$^{-2}$\,ksec$^{-1}$\,\AA$^{-1}$. The 2021d37.1
spectrum (grey shadings) is compared to various 2006 and 2021 spectra as
marked in the top legends. The top panel shows the global picture
logarithmically and with a larger wavelength range while the panels below
focus into narrower regions of interest with line labels, label
colours representing different ions. Unidentified features are marked with
numerical wavelength values.
In the bottom two panels, the \chandra/LETGS spectrum taken on day 2006d39.7 is
included in blue showing the excess above a blackbody fit, shifted up by
20\,ph\,cm$^{-2}$\,ksec$^{-1}$\,\AA$^{-1}$.
}
\end{figure*}

\subsection{Spectra}
\label{sect:results:spectra}

\subsubsection{EPIC spectrum compared to RGS}
\label{sect:results:spectra:epic}

As reported by \cite{page2022}, the first glimpse of a soft 'bump' was seen in \swift/XRT spectra on day 2021d20.6 while this 'bump' increased significantly on day 2021d26.1 \citep{sssstart_2021}. In spectra with the resolution of the \swift/XRT or \xmm/EPIC, such 'bumps' appear as smooth peaks with the shape of a blackbody, see e.g., figure~4 in \cite{swnovae}. As a test whether such a soft 'bump' can reliably be interpreted as early onset of the SSS phase, we inspect the simultaneous low-resolution EPIC/MOS1 and RGS1 spectra of RS\,Oph taken on day 2006d26.1\footnote{the same exercise is not possible with the day 2021d37.1 observation because there are no MOS data and pn only reaches down to 0.7\,keV}, shown in Fig.~\ref{fig:epic_rgs}. The APEC model represents the short-wavelength component arising from shock interactions between the nova ejecta and the stellar wind of the companion star while the blackbody model represents the soft 'bump'. Indeed, the soft 'bump' in the MOS1 spectrum is reproduced well with the blackbody component, and if only CCD-resolution spectra were available, we would conclude with high certainty that we are seeing early SSS continuum emission. However, the simultaneous RGS1 spectrum shown in the bottom panel shows good agreement with the APEC model but no agreement at all with the blackbody model.\\

The striking contrast between the successful blackbody fit to the low-resolution spectrum and its failure to reproduce the contemporaneous high-resolution data demonstrates that critical spectral information is lost at low resolution. We made the experiment to degrade the RGS spectrum by applying a gaussfilter with $\sigma=4$\,\AA\ and indeed obtain a blackbody spectrum.

It is highly likely that the early 'bumps' in the \swift/XRT spectra can originate from similarly complex soft emission, and the fact that the day 2021d37.1 RGS spectrum looks similar (see figure 6 in \citealt{ness2023} and the next sections in this work), such a complex early soft component may be rather common.

\subsubsection{Comparison of grating spectra from different epochs}
\label{sect:results:spectra:rgs}

The RGS spectrum obtained on day 2021d37.1 is shown in Fig.~\ref{fig:xmmspec37} (grey shading, photon flux units) together with selected RS\,Oph grating spectra. At wavelengths $\lambda \lesssim 18$\,\AA, all spectra are remarkably similar and trace the shock emission discussed by \cite{page2022,orio2022}. In contrast, at $\lambda \gtrsim 18$\,\AA, day 2021d37.1 exhibits a complex soft emission component that was absent on day 2021d21.1 (black) and resembles, although at higher flux, the transient soft component observed on day 2006d26.1 (red; \citealt{nelson07,rsophshock}). While this component could represent the onset of SSS emission, its appearance is dominated by emission lines rather than the smooth continuum typical of SSS spectra. As noted by \cite{basi,nessobsc}, it was transient and concentrated between the strong N\,{\sc vii} and N\,{\sc vi} lines at 24.78 and 28.78\,\AA, respectively.\\

The second panel highlights the evolution of the shock emission. In both the 2006 and 2021 outbursts, the shock plasma follows the same cooling sequence: the continuum progressively softens, and lines from highly ionised species (e.g. Si\,{\sc xiv} and Mg\,{\sc xii}) fade more rapidly than lower-ionisation lines such as Ne\,{\sc x}. The spectra from both eruptions align closely along this trend, indicating essentially identical shock evolution \citep{page2022,orio2022}.\\

While at short wavelengths, the day 2006d26.1 spectrum fits in nicely with the other shock spectra, e.g. on day 2021d21.1, the \chandra/LETGS spectrum taken on day 2006d39.7 was already dominated by bright SSS continuum \citep{nessrsoph} at $\lambda<15$, thus distinctly different from the same-episode spectrum on day 2021d37.1.
To compare the weak line emission on top of the bright SSS continuum spectrum seen on day 2006d39.7 with day 2021d37.1, we subtracted a best-fit blackbody model from the day 2006d39.7 spectrum and show the residual emission as the blue histogram in the lower panels of Fig.~\ref{fig:xmmspec37}. A constant offset of 20\,ph\,cm$^{-2}$\,ksec$^{-1}$\,\AA$^{-1}$ was added to display both emission and absorption features clearly. Without applying any scaling, the residual emission features, as well as an absorption feature at 26.8\,\AA, are found to be of comparable strength on days 2006d39.7 and 2021d37.1.\\

The similarity of the excess line emission in 2006d39.7 and 2021d37.1 suggests that comparable plasma conditions were present during these two otherwise spectrally distinct phases.\\

Line labels are included with different colours for different elements. While all features in the shock spectra (second panel) can clearly be identified, several features in the soft spectra remain unidentified (see many question marks in the bottom panel). 
The notation using $\infty$, e.g., N\,{\sc vii}$\infty$, refers to the ionisation threshold of the respective ion. These energies can mark either absorption edges in continuum spectra or Radiative Recombination Continuum (RRC) features. RRCs are produced when free electrons recombine with highly ionised ions and emit the excess kinetic energy as radiation. They appear as edge-like emission features with a sharp onset at the ionisation threshold energy. The width of an RRC is directly related to the electron temperature of the recombining plasma, with narrow RRCs indicating very cold electrons. The widths of features marked in the second and third panels as N\,{\sc vii}$\infty$ and N\,{\sc vi}$\infty$ correspond to an electron temperature of 1-2\,eV.

\subsection{Spectral Variations on day 2021d37.1}
\label{sect:results:specvar}

\begin{figure*}[!ht]
\resizebox{\hsize}{!}{\includegraphics{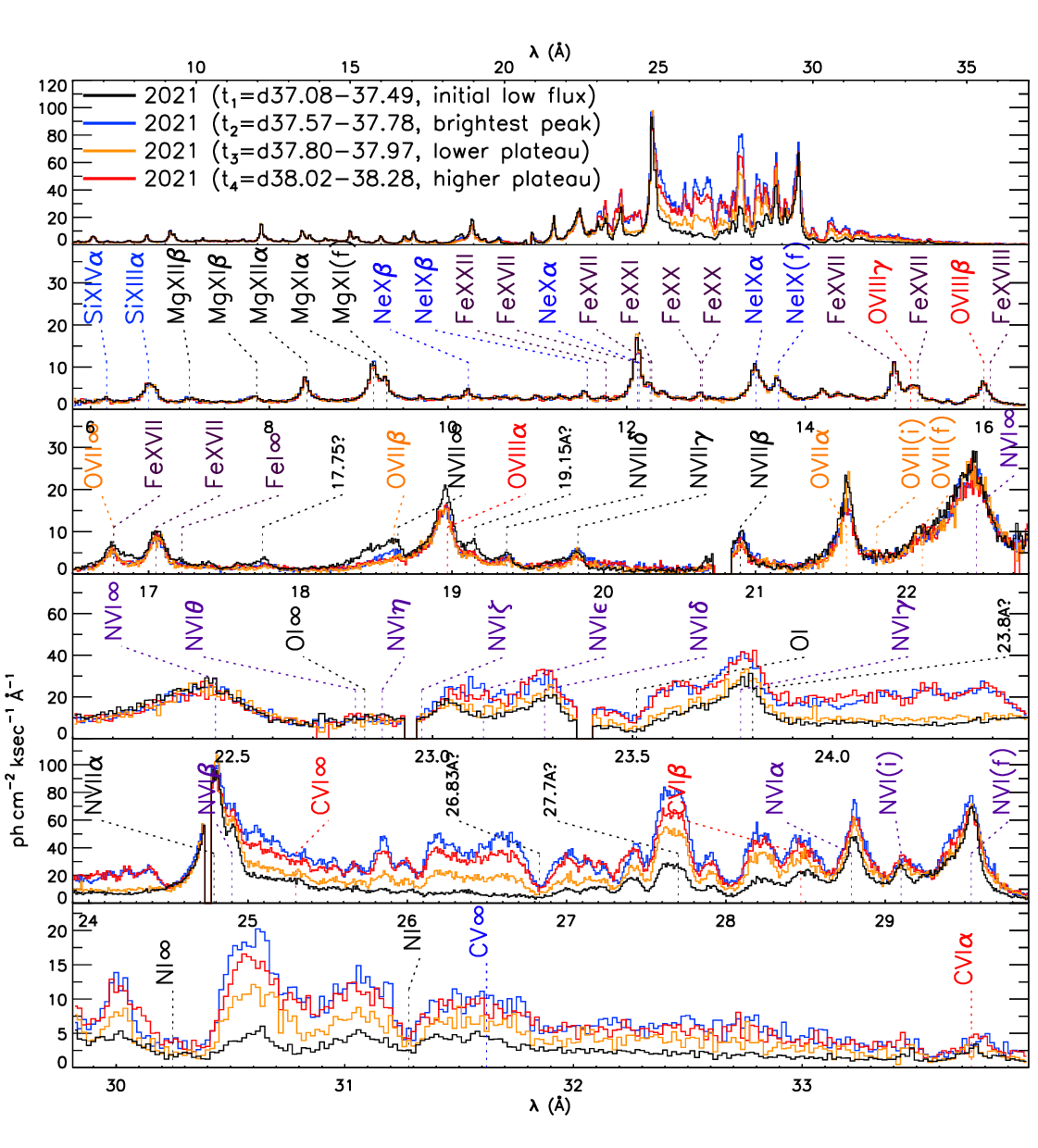}}
 \caption{\label{fig:cmp37}Comparison of the spectra extracted from the four time intervals marked with the same colours in the left graph of Fig.~\ref{fig:xmmlc}. Line labels are included at rest wavelengths using different colours for different ions. Some features are marked with their wavelength values if no clear identification is found.}
\end{figure*}

\begin{figure}[!ht]
\resizebox{\hsize}{!}{\includegraphics{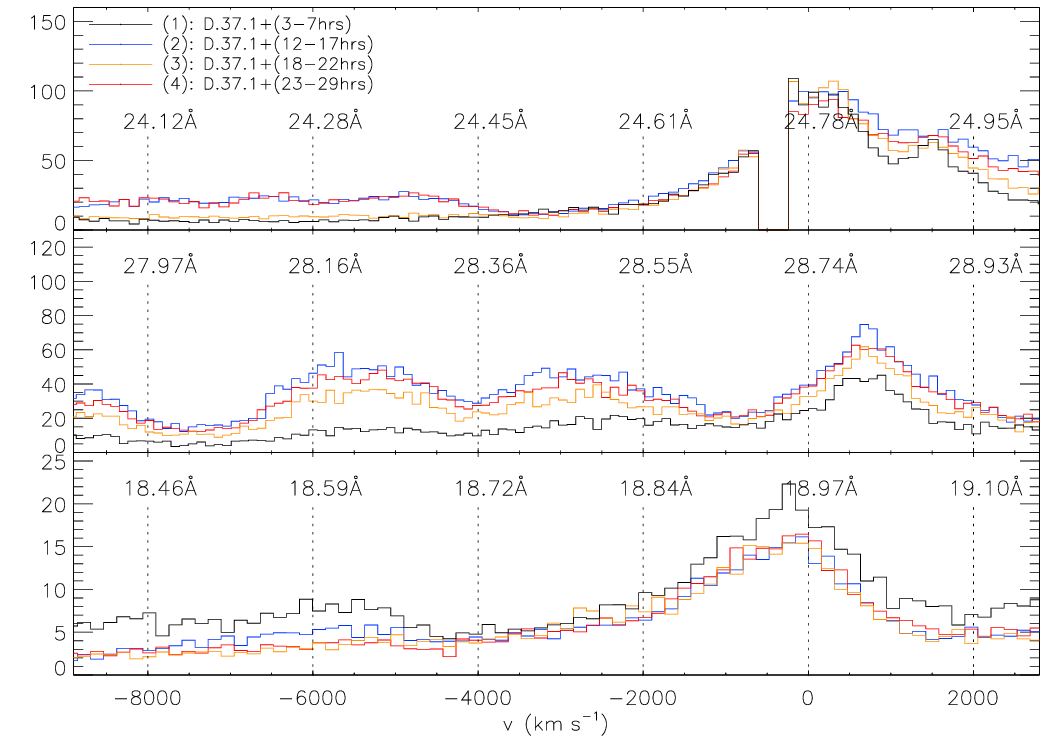}}
 \caption{\label{fig:lprof}Comparison of line profiles for the four times intervals defined in Fig.~\ref{fig:xmmlc} in velocity space for the lines of N\,{\sc vii} ($\lambda_0=24.78$\,\AA), N\,{\sc vi} ($\lambda_0=28.78$\,\AA), and O\,{\sc viii} ($\lambda_0=18.97$\,\AA), from top to bottom, respectively. Vertical dashed lines mark rest wavelengths serving for identification of features that may not be blue-/red-shifted lines.}
\end{figure}

In Fig.~\ref{fig:cmp37}, RGS spectra extracted from the four time intervals defined in Fig.~\ref{fig:xmmlc} are compared. 
In the short-wavelength range $<16.5$\,\AA\ no spectral variations are present. 
The initial low-flux interval (t$_1$, black) is generally the faintest spectrum but shows several emission features that subsequently weaken despite the rise in source brightness, including structures near 16.9\,\AA, 18.6\,\AA, 18.97\,\AA\ (O\,{\sc viii}), 19.15\,\AA, and 21.6\,\AA\ (O\,{\sc vii}).
The features near 16.9\,\AA\ and 18.6\,\AA\ coincide with the O\,{\sc vii} and N\,{\sc vii} ionisation energies, while the 19.15\,\AA\ feature might originate from O\,{\sc vii} di-electronic recombination lines. A possible, albeit at best marginal, decrease of the O\,{\sc vii} forbidden line at 22.1\,\AA\ follows the same trend and, if real, would disfavour a self-absorption scenario. The selective fading of emission features between 18 and 23\,\AA\ without a corresponding continuum change remains difficult to explain.

At wavelengths above 23\,\AA\ the source exhibits much stronger variability, with enhanced emission during the later bright intervals, except during the second low-flux episode (t$_3$), which resembles t$_1$ up to about 25\,\AA. The transition near 23\,\AA\ may be connected to ionisation of neutral oxygen ($E>543$\,eV), although simple changes in neutral O absorption cannot directly explain the observed behaviour.

Line-profile comparisons (Fig.~\ref{fig:lprof}) show that the H-like N\,{\sc vii} and O\,{\sc viii} lines remain close to their rest wavelengths, whereas the He-like N\,{\sc vi} line appears redshifted by approximately $\sim 650$\,km\,s$^{-1}$, possibly due to absorption affecting the blue wing of the line.

\begin{figure*}[!ht]
\resizebox{\hsize}{!}{\includegraphics{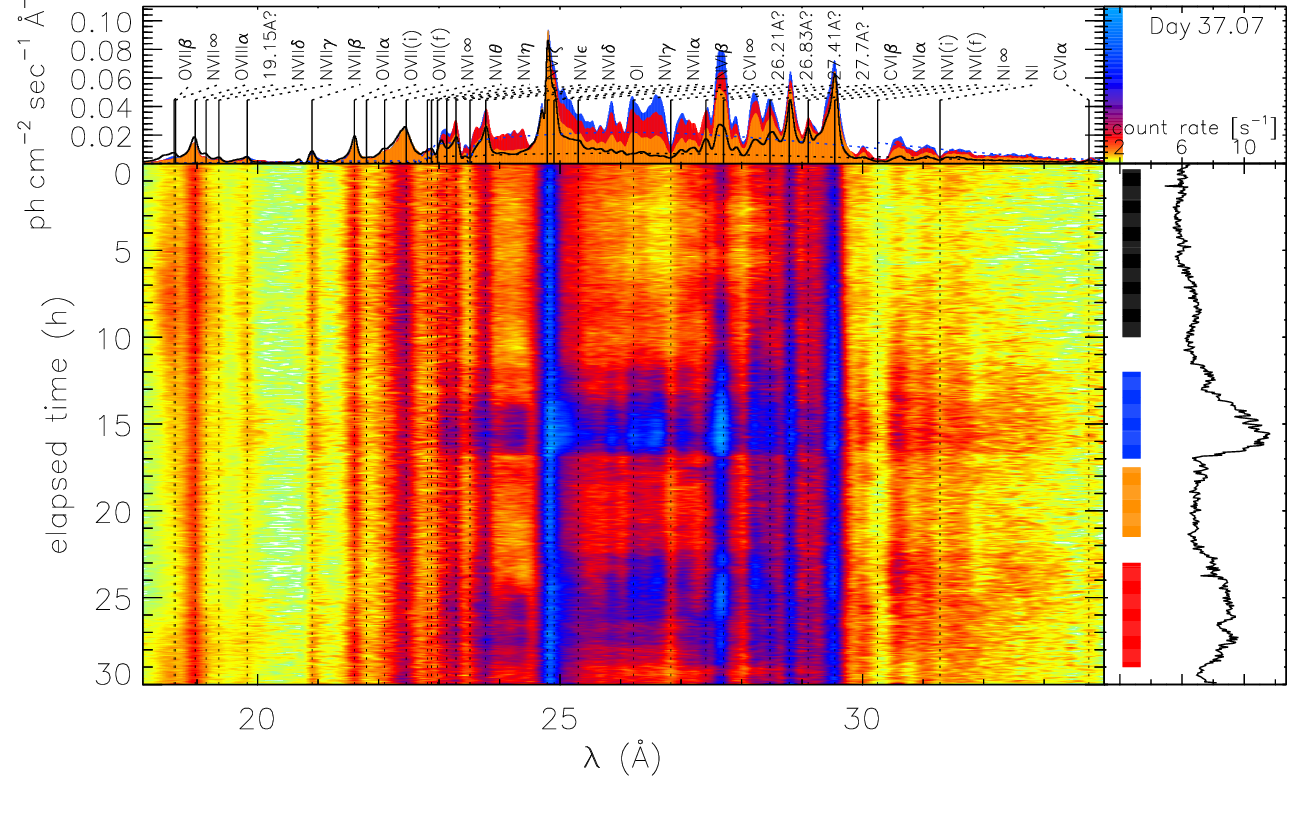}}
\caption{\label{fig:smap37}Spectral Time map based on 540 RGS spectra
extracted from adjacent 200-s time intervals in the 2021d37.1 observation. The main panel shows the
spectra with wavelength across, time down, and colours representing flux
following the (non-linear) bar in the top right panel along the vertical
flux axis. The shadings in the bottom right panel indicate time intervals
from which the spectra shown in the same colour shadings in the top left
panel were extracted.}
\end{figure*}

The spectral time map (Fig.~\ref{fig:smap37}) reveals that the transient excess around 18.5\,\AA\ is present only during the initial faint phase, disappears as the long-wavelength soft emission rises, and reappears near the end of the observation after the 25-29\,\AA\ emission weakens. Similar but weaker behaviour is seen near 22.5\,\AA, features, tentatively identified as the N\,{\sc vi} RRC feature. This common evolution favours interpreting the 18.5\,\AA\ feature as the N\,{\sc vii} RRC rather than an O\,{\sc vii} 1s-3p transition. The strongest and fastest variability originates from the highly structured emission complex between 25.5 and 27.5\,\AA, whereas several other features remain stable or evolve on much longer timescales.

\subsection{Comparison with other SSS spectra}

\begin{figure*}[!ht]
\resizebox{\hsize}{!}{\includegraphics{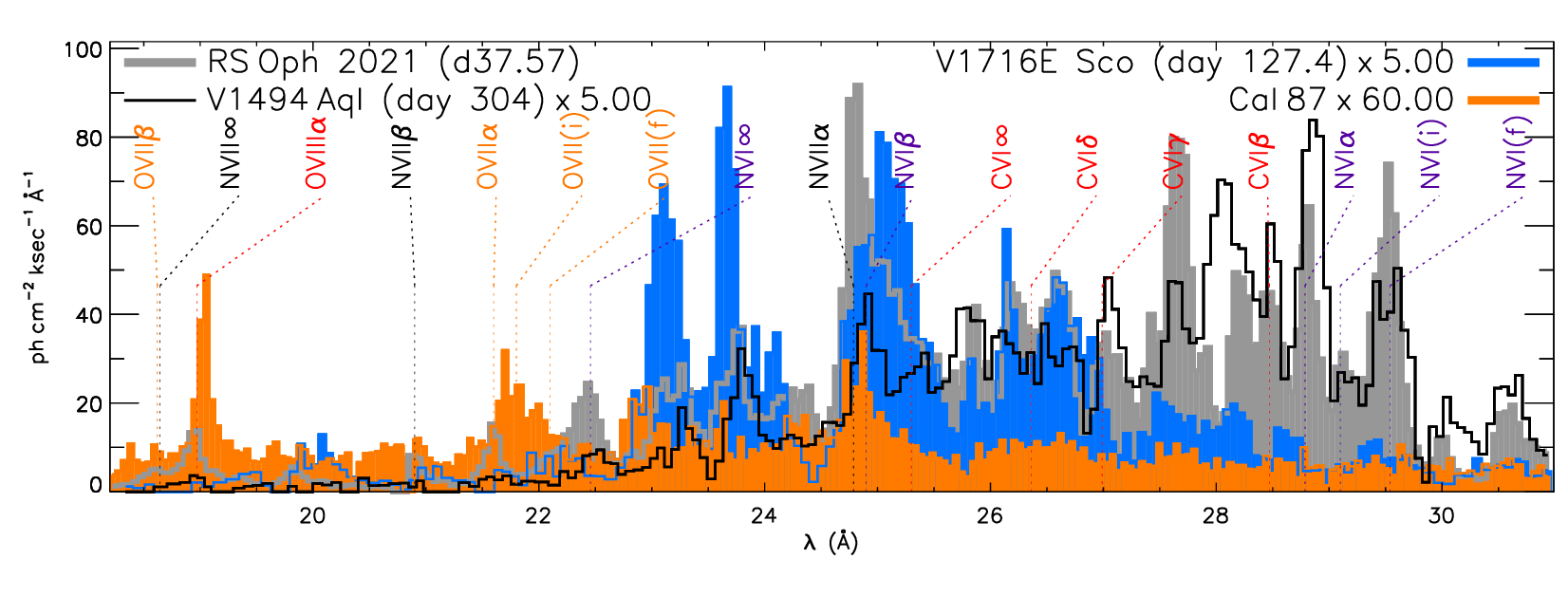}}
\caption{\label{fig:cmpnovae}Comparison of the new RGS spectrum of RS\,Oph
 on day 2021d37 ($t_2$ interval in Fig.~\ref{fig:xmmlc}, grey) with
 the persistent high-inclination angle SSS Cal\,87 in the LMC
 (orange shadings, upscaled
 by a factor 60) and the novae V1716\,Sco (on day 127.4 after discovery,
 blue shadings, upscaled by factor 5) and V1494\,Aql (304 days after
 visual maximum, black historgram, upscaled by factor 5).
 }
\end{figure*}

In pursuit of the question whether we are dealing with a unique phenomenon, we inspected
several other X-ray grating SSS spectra and show in Fig.~\ref{fig:cmpnovae} comparison of
the brightest sub-spectrum from the day2021d37 observation (time interval $t_2$) with
those cases that best support the identification of suitable and unsuitable future analysis
approaches in \S\ref{sect:perspectives}.
First, we compare our reference spectrum of the persistent SSS Cal\,87 (orange shadings) that
is the clearest case of an SSe spectrum according to the \cite{nessobsc} classification and see
no resemblance in these two spectra.
\cite{nessobsc} not only reported clear SSe and SSa spectra but also found a number of intermediate cases such as HV\,Cet, V5116\,Sgr, and V1494\,Aql (see their comparison figure~4), and we include the spectrum of V1494\,Aql (see \citealt{rohrbach09}) in our comparison plot Fig.~\ref{fig:cmpnovae} with the black histogram where it can be seen that this spectrum shows remarkable resemblance with our spectrum. The more recent SSS spectrum of V1716\,Sco (blue shadings) could also be seen to be an intermediate case between SSa and SSe and shows some resemblance with our reference spectrum of RS\,Oph.\\

These comparisons only concern overall spectral morphology. Differences in ionisation conditions and other plasma parameters preclude using the resemblance or lack of resemblance by itself to establish a common or different emission mechanism.

\section{Perspectives for physical interpretations}
\label{sect:perspectives}

In \S\ref{sect:results}, a detailed account of the spectral and timing properties is presented with the following key results:
\begin{itemize}
\item The ordinary shock emission and the long-term UV brightness evolved continuously on time scales of days where the shock X-rays evolved very similarly in the 2006 and 2021 outbursts;
\item Superposed on this shock emission is a highly variable soft component that appears like a blackbody in low resolution, whereas the high resolution data show that it is intrinsically highly line-dominated;
\item Not all spectral features followed the same brightness trends;
\item The temporal behaviour of the soft component is inconsistent with a simple increase or decrease of continuum emission or absorption alone;
\item Some of the features may tentatively be identified as Radiative Recombination Continuum (RRC) features;
\item The peculiar soft spectrum on day 2021d37.1 strongly resembles the residual line emission above the bright SSS continuum on day 2006d39.7.
\item The soft component shows some qualitative resemblance with the nova spectra of V1494\,Aql and V1716\,Sco but not the persistent SSS Cal 87.
\end{itemize}

Firstly, we would like to stress that, while this complex type of soft spectrum is distinctly different from typical SSS emission, there are three pieces of evidence suggesting that a highly obscured SSS spectrum might influence the formation of the observed complex spectrum:
\begin{itemize}
\item The [Fe\,{\sc x}] flux evolution indicates that ionising
SSS emission was already present around day 23 after optical peak \citep{page2022},
\item Detection of a 35s period \citep{nelson07},
\item The low-resolution EPIC spectra form a perfect blackbody shape (this work).
\end{itemize}
How exactly, a deeply underlying SSS emission could influence the observable spectrum if processes dominate that produce other than photospheric emission needs to be determined with quantitative models.
Photoionisation models are a natural starting point. However, a disagreement between the observations and a simplified implementation would not necessarily demonstrate that photoionisation is unimportant, because the result may depend on assumptions regarding geometry, density structure, absorption, dynamics, clumpiness, or other departures from equilibrium.
Establishing whether a discrepancy reflects a limitation of a particular model realisation, an inadequate choice of parameters, or a shortcoming of the underlying physical scenario would itself require a dedicated investigation. We therefore regard such modelling as an important but separate research effort and focus here on establishing the observational constraints that future modelling should aim to reproduce.\\

The comparable excess line emission above the bright SSS continuum on day 2006d39.7 and in the day 2021d37.1 spectrum, without applying a flux scaling in Fig.~\ref{fig:xmmspec37}, suggests that this component can coexist with bright SSS emission and may originate in a physically distinct region.
One possible explanation is that the plasma producing the observed spectrum is exposed to a stronger ionising flux than is directly visible along our line of sight.\\

The combined spectral and temporal behaviour suggests that the transient soft component is not an extension of the shock emission, but originates at least partially in a distinct recombining plasma that may be closely linked to the emergence of the SSS phase and may represent the same emission component seen above the SSS continuum during the 2006 outburst.
The qualitative similarities between the 2006d26.1 and 2021d37.1 spectra indicate that RS Oph repeatedly develops the same transient soft, line-dominated emission component shortly before the emergence of the supersoft-source phase and may not be disturbed by bright SSS emission. A common physical mechanism may be operating in both outbursts.\\

The presence of emission lines in SSS spectra has been seen in several novae especially in the SSe type spectra defined by \cite{nessobsc}. 
While the later bright SSS spectra of RS Oph belong to the SSa class, temporary blocking of central SSS emission could lead to observing an SSe spectrum for a limited time, similar to V5116\,Sgr that \cite{nessobsc} showed to transition between SSa and SSe spectra. The comparison of the SSe spectrum of Cal 87 with our RS\,Oph spectrum on day 2021d37.1 in Fig.~\ref{fig:cmpnovae} shows no resemblance, and these particular emission lines do not qualify to classify this spectrum as an SSe spectrum. On the other hand, the resemblance with the intermediate cases V1494\,Aql and V1716\,Sco suggests that the day 2006d26.1 and 2021d37.1 spectra of R\,Oph belong to this intermediate group. In particular, the observation of an SSe spectrum does not necessarily mean a high-inclination system as it does not necessarily need an accretion disk to block bright SSS continuum emission.\\

The complexity of the spectral features, their time evolution, and the uncertain influence of photoionising continuum radiation demonstrate that the selection of physically meaningful models is not straightforward. The appearance of features tentatively identified as RRC lines, together with the strong variability of individual emission complexes indicate that recombination processes play an important role in the origin of the transient soft emission. A natural interpretation would be photoionisation by an underlying SSS source, consistent with the evidence from the [Fe\,{\sc x}] evolution and the 35-s modulation discussed above. However, the observations do not suggest a simple, steady-state photoionised plasma. The rapid spectral variability and the evolving contribution of individual features instead point to a dynamic environment in which multiple plasma components and physical processes may coexist.\\

The features tentatively identified as RRC lines may provide an important clue. Narrow RRCs indicate recombination of highly ionised ions with relatively cold electrons and therefore require close interaction between hot and cool plasma components. Such conditions may naturally occur at interfaces between shocked nova ejecta and cooler surrounding material. In recent years, RRCs together with enhanced high-$n$ transitions have been used as diagnostics of charge-exchange and plasma-mixing processes in nova ejecta \citep{mitrani2024,mitrani_v1716sco}. These studies demonstrate that structured soft X-ray spectra containing prominent recombination features do not uniquely imply photoionisation and may instead arise from interactions between hot ionised plasma and cold gas.
We do not unambiguously identify all CXC-specific diagnostics, such as a quantitatively established enhancement of high-$n$ transitions, in the present RS\,Oph spectra; consequently, based on the observations alone, we regard charge exchange as a mechanism warranting future investigation rather than as a demonstrated interpretation.\\

Since the observed emission must originate in regions that are sufficiently optically thin for the radiation to escape, it may arise at the interface between hot ejecta and cooler surrounding material. Charge-exchange processes can occur naturally in such environments when electrons are transferred from neutral or weakly ionised atoms into excited states of highly ionised species. This scenario was successfully applied by \cite{mitrani2024} to YZ\,Ret, where the observed spectrum was interpreted as evidence for mixing between hot shocked plasma and cold gas. Although the spectrum of YZ\,Ret differs substantially from that of RS\,Oph, the qualitative resemblance between the RS\,Oph spectra and the nova V1716\,Sco suggests that the modelling approach of \cite{mitrani_v1716sco} may provide one possible framework for quantitative investigation of the transient soft emission observed here. Motivated by this similarity, the same methodology has recently been applied to the RS\,Oph spectra discussed in this work in a study that is currently under review\footnote{Preprint available at \url{https://www.researchsquare.com/article/rs-9899519/v1}}.\\

The resemblance of RS\,Oph to V1716\,Sco, and to a lesser extent V1494\,Aql, raises the possibility that charge-exchange processes contribute to at least some of the intermediate spectra lying between the canonical SSa and SSe classes identified by \cite{nessobsc}. Future quantitative modelling will be required to determine whether this similarity reflects a common physical origin.

\section{Summary and Conclusions}
\label{sect:concl}

\begin{itemize}
\item We analysed the soft X-ray spectra obtained during the 2006 and 2021 outbursts of RS\,Oph at epochs preceding the appearance of the classical bright SSS spectra observed later in both eruptions.
\item Despite occurring at different times after optical maximum, the spectra obtained on 2006d26.1 and 2021d37.1 display remarkably similar high-resolution spectral morphology, indicating that a common transient emission component was present during both outbursts.
\item While the corresponding EPIC spectra are consistent with a smooth blackbody-like continuum, the RGS spectra are highly structured and contain numerous emission features, some of which may be associated with radiative recombination continua yielding electron temperatures 1-2\,eV.
\item Independent evidence from the evolution of the [Fe\,{\sc x}] coronal line flux and the detection of the 35-s oscillation suggests that ionising SSS emission was already present while not yet directly observable along our line of sight.
\item The observed spectra do not resemble classical emission-dominated SSe spectra such as Cal\,87, but show similarities to intermediate objects discussed by \cite{nessobsc}, including V1494\,Aql and V1716\,Sco.
\item The presence of probable recombination features together with the spectral resemblance to V1716\,Sco indicates that physical scenarios involving interactions between highly ionised plasma and cooler material, including recently proposed charge-exchange models, provide one possible framework, warrant quantitative investigation.
\item Quantitative modelling will be required to determine the relative importance of photoionisation, recombination, charge exchange, and ejecta-wind interactions in producing the transient soft X-ray emission during the emergence of the SSS phase in RS\,Oph.
\end{itemize}

Our results suggest that the emergence of observable SSS emission in RS\,Oph is accompanied by a distinct transient soft X-ray component whose properties cannot be explained solely by the appearance of a white-dwarf atmosphere spectrum. The recurrence of this phenomenon in both the 2006 and 2021 eruptions indicates that it is
a recurring feature of the observed RS Oph outburst evolution and provides a valuable laboratory for studying the interaction between nuclear-burning emission, shocked ejecta, and surrounding circumstellar material.

\begin{acknowledgements}
S. Starrfield gratefully acknowledges partial support from NASA grants to ASU.
We are grateful to the late Prof. Dame Carole Jordan (1941-2026) for extensive discussions of the unusual spectral features observed during the onset of the supersoft-source phase in RS Oph. Her examination of atomic data and line lists explored possible identifications involving highly ionised argon, calcium, and satellite lines associated with H-like and He-like nitrogen. While no unique identifications could be established, her analysis highlighted the potential importance of weak transitions, non-equilibrium effects, and radiative-transfer processes in shaping these spectra. Her insights contributed significantly to our understanding of the complexity of the observational material presented here.
\end{acknowledgements}

\bibliographystyle{aa}
\bibliography{mybib}

\appendix

\section{Appendix}
\label{appendix:g7005OOL}

During ObsID 0841930501, an anomaly called "G7005 OOL" occurred for the RGS1, and a standard {\tt rgsproc} run does not produce any events.
The procedure described here has been provided by the \xmm\ SOC to process the currently available data. A correction is being implemented by the SOC in the SAS, and the ODF will be fixed so that the data can be processed without this procedure.

This anomaly requires the housekeeping good-time-interval file {\tt hkgti} to be produced with non-standard parameters 'G7005 G5063'. The modified {\tt hkgti} file then needs to be integrated in the {\tt rgsproc} chain. A full {\tt rgsproc} run produces a default hkgti file, overwriting any existing hkgti file, and it thus needs to be interrupted using the {\tt finalstage} and {\tt entrystage options}. We thus processed the RGS1 data as follows:

\begin{itemize}

\item Produce the {\tt hkgti} file with the command:\\
{\tt hkgtigen instrument=rgs1 gtiset=P0841930501R1S002hkgti\_0000.FIT diagfile=STDOUT plotgtis=no diagnosticmode=no withgtiset=yes parameters='G7005 G5063' withparameters=yes except=yes overrideparameters='' overrideparametervalues='' withoverrideparameters=no timecolumn=TIME prefraction=0.5 postfraction=0.5  -w 1 -V 6}

\item Take a backup of the thus created housekeeping gti file:\\
{\tt mv P0841930501R1S002hkgti\_0000.FIT P0841930501R1S002hkgti\_0000.FIT.sav}

\item 
Run {\tt rgsproc} only until it produces the default {\tt hkgti} file:\\
{\tt rgsproc finalstage=1:events withinstexpids=yes instexpids='R1S002'}

\item Overwrite the default house keeping file produced during this run with the backup:\\
{\tt mv P0841930501R1S002hkgti\_0000.FIT.sav P0841930501R1S002hkgti\_0000.FIT}

\item The remaining {\tt rgsproc} chain are then executed starting from the stage {\tt angles} with\\
{\tt rgsproc  entrystage=2:angles withinstexpids=yes instexpids='R1S002'}

\item Create RGS1 light curve with\\
{\tt evselect table=P0841930501R1S002EVENLI0000.FIT:EVENTS energycolumn=PI expression='REGION(P0841930501R1S002SRCLI\_0000.FIT:RGS1\_SRC1\_SPATIAL,M\_LAMBDA,XDSP\_CORR)' withrateset=yes rateset="R1\_source\_lightcurve\_raw.lc" timebinsize=10 maketimecolumn=yes makeratecolumn=yes}

\item Create RGS1 fluxed spectrum with\\
{\tt rgsfluxer bkg='' pha='P0841930501R1S002SRSPEC1001.FIT' rmf='P0841930501R1S002RSPMAT1001.FIT'}

\item Create RGS2 products with\\
{\tt rgsproc withinstexpids=yes instexpids='R2S004'}

\item Create combined RGS1+RGS2 1st+2nd order fluxed spectrum with\\
{\tt rgsfluxer bkg='' pha='P0841930501R1S002SRSPEC1001.FIT P0841930501R1S002SRSPEC2001.FIT P0841930501R2S004SRSPEC1001.FIT P0841930501R2S004SRSPEC2001.FIT' rmf='P0841930501R1S002RSPMAT1001.FIT P0841930501R1S002RSPMAT2001.FIT P0841930501R2S004RSPMAT1001.FIT P0841930501R2S004RSPMAT2001.FIT'}

\end{itemize}

\end{document}

%% file: logtab.tex
0841930501 & pn & Timing/Burst & Thick & 2021-09-15T15:07:26 & 2021-09-16T20:33:09 & 105.83\,ks\\
day 37.07$^a$ & RGS1 & \multicolumn{2}{l}{RGS Small Window}  & 2021-09-15T14:38:15 & 2021-09-16T20:37:03 & 107.82\,ks\\
 & RGS2 & \multicolumn{2}{l}{RGS Small Window} & 2021-09-15T14:39:04 & 2021-09-16T20:35:48 & 107.72\,ks\\
 & OM & IMAGE+FAST & UVW2 & 2021-09-15T14:47:30 & 2021-09-15T16:00:50 & 4400\,s\\
 & OM & IMAGE+FAST & UVW2 & 2021-09-15T16:05:58 & 2021-09-15T17:19:16 & 4400\,s\\
 & OM & IMAGE+FAST & UVW2 & 2021-09-15T17:24:25 & 2021-09-15T18:37:43 & 4400\,s\\
 & OM & IMAGE+FAST & UVW2 & 2021-09-15T18:57:53 & 2021-09-15T20:11:11 & 4400\,s\\
 & OM & IMAGE+FAST & UVW2 & 2021-09-15T20:16:18 & 2021-09-15T21:29:38 & 4400\,s\\
 & OM & IMAGE+FAST & UVW2 & 2021-09-15T21:34:46 & 2021-09-15T22:48:07 & 4400\,s\\
 & OM & IMAGE+FAST & UVW2 & 2021-09-15T22:53:13 & 2021-09-16T00:06:33 & 4400\,s\\
 & OM & IMAGE+FAST & UVW2 & 2021-09-16T00:11:40 & 2021-09-16T01:25:00 & 4400\,s\\
 & OM & IMAGE+FAST & UVW2 & 2021-09-16T01:30:07 & 2021-09-16T02:43:28 & 4400\,s\\
 & OM & IMAGE+FAST & UVW2 & 2021-09-16T02:48:35 & 2021-09-16T04:01:54 & 4400\,s\\
 & OM & IMAGE+FAST & UVW2 & 2021-09-16T04:07:00 & 2021-09-16T05:20:20 & 4400\,s\\
 & OM & IMAGE+FAST & UVW2 & 2021-09-16T05:25:28 & 2021-09-16T06:38:48 & 4400\,s\\
 & OM & IMAGE+FAST & UVW2 & 2021-09-16T06:43:55 & 2021-09-16T07:57:15 & 4400\,s\\
 & OM & IMAGE+FAST & UVW2 & 2021-09-16T08:22:22 & 2021-09-16T09:35:42 & 4400\,s\\
 & OM & IMAGE+FAST & UVW2 & 2021-09-16T09:40:48 & 2021-09-16T10:54:09 & 4400\,s\\
 & OM & IMAGE+FAST & UVW2 & 2021-09-16T10:59:15 & 2021-09-16T12:12:35 & 4400\,s\\
 & OM & IMAGE+FAST & UVW2 & 2021-09-16T12:17:43 & 2021-09-16T13:31:04 & 4400\,s\\
 & OM & IMAGE+FAST & UVW2 & 2021-09-16T13:36:11 & 2021-09-16T14:49:30 & 4400\,s\\
 & OM & IMAGE+FAST & UVW2 & 2021-09-16T14:54:37 & 2021-09-16T16:07:57 & 4400\,s\\
 & OM & IMAGE+FAST & UVW2 & 2021-09-16T16:13:04 & 2021-09-16T17:26:24 & 4400\,s\\
 & OM & IMAGE+FAST & UVW2 & 2021-09-16T17:31:32 & 2021-09-16T18:44:50 & 4400\,s\\
 & OM & IMAGE+FAST & UVW2 & 2021-09-16T18:49:58 & 2021-09-16T20:03:17 & 4400\,s\\
\hline
0410180201  & RGS1 & \multicolumn{2}{l}{Spectroscopy} & 2006-03-10T23:02:51 & 2006-03-11T02:20:57 & 11.78\,ks\\
day 26.12$^b$  & RGS2 & \multicolumn{2}{l}{Spectroscopy} & 2006-03-10T23:02:56 & 2006-03-11T02:20:57 & 11.78\,ks\\
0410180301  & RGS1 & \multicolumn{2}{l}{Spectroscopy} & 2006-04-07T21:04:29 & 2006-04-08T02:20:31 & 13.87\,ks\\
day 54.04$^b$& RGS2 & \multicolumn{2}{l}{Spectroscopy} & 2006-04-07T21:04:34 & 2006-04-08T02:20:31 & 13.87\,ks\\
7296 (day 39.7$^b$) & \chandra & LETGS & HRC-S & 2006-03-24T12:25:22 & 2006-03-24T15:38:20 & 9.97\,ks\\
\hline
0072$^c$ & \chandra & LETGS & HRC-S & 2000-10-01T10:08:56 & 2000-10-01T15:39:30 & 19.8\,ks\\ 
28048$^d$ & \chandra & LETGS & HRC-S & 2023-08-26T01:35:32 & 2023-08-26T06:00:51 & 15.8\,ks\\
0153250101$^e$ & RGS & \multicolumn{2}{l}{Spectroscopy} & 2003-04-18T20:54:17 & 2003-04-19T18:44:31 & 78.6\,ks\\

%% file: mybib.bib
@ARTICLE{ole,
author = {{K{\"o}nig}, Ole and {Wilms}, J{\"o}rn and {Arcodia}, Riccardo and {Dauser}, Thomas and {Dennerl}, Konrad and {Doroshenko}, Victor and {Haberl}, Frank and {H{\"a}mmerich}, Steven and {Kirsch}, Christian and {Kreykenbohm}, Ingo and {Lorenz}, Maximilian and {Malyali}, Adam and {Merloni}, Andrea and {Rau}, Arne and {Rauch}, Thomas and {Sala}, Gloria and {Schwope}, Axel and {Suleimanov}, Valery and {Weber}, Philipp and {Werner}, Klaus},
title = "{X-ray detection of a nova in the fireball phase}",
journal = {Nature},
year = 2022,
month = may,
volume = {605},
number = {7909},
pages = {248-250},
doi = {10.1038/s41586-022-04635-y},
adsurl = {https://ui.adsabs.harvard.edu/abs/2022Natur.605..248K}
}

@ARTICLE{heuvel,
author = {{van den Heuvel}, E.~P.~J. and {Bhattacharya}, D. and {Nomoto}, K. and
{Rappaport}, S.~A.},
title = "{Accreting white dwarf models for CAL 83, CAL 87 and other ultrasoft X-ray sources in the LMC}",
journal = {A\&A},
year = 1992,
month = aug,
volume = 262,
pages = {97-105}
}

@ARTICLE{osborne11,
author = {{Osborne}, J.~P. and {Page}, K.~L. and {Beardmore}, A.~P. and
{Bode}, M.~F. and {Goad}, M.~R. and {O'Brien}, T.~J. and {Starrfield}, S. and
{Rauch}, T. and {Ness}, {J.-U.} and {Krautter}, J. and {Schwarz}, G. and
{Burrows}, D.~N. and {Gehrels}, N. and {Drake}, J.~J. and {Evans}, A. and
{Eyres}, S.~P.~S.},
title = "{The Supersoft X-ray Phase of Nova RS Ophiuchi 2006}",
journal = {ApJ},
archivePrefix = "arXiv",
eprint = {1011.5327},
primaryClass = "astro-ph.HE",
year = 2011,
month = feb,
volume = 727,
pages = {124},
doi = {10.1088/0004-637X/727/2/124}
}

@ARTICLE{orio23,
author = {{Orio}, Marina and {Gendreau}, Keith and {Giese}, Morgan and {Luna}, Gerardo Juan M. and {Magdolen}, Jozef and {Strohmayer}, Tod E. and {Zhang}, Andy E. and {Altamirano}, Diego and {Dobrotka}, Andrej and {Enoto}, Teruaki and {Ferrara}, Elizabeth C. and {Ignace}, Richard and {Heinz}, Sebastian and {Markwardt}, Craig and {Nichols}, Joy S. and {Parker}, Michael L. and {Pasham}, Dheeraj R. and {Pei}, Songpeng and {Pradhan}, Pragati and {Remillard}, Ron and {Steiner}, James F. and {Tombesi}, Francesco},
title = "{The RS Oph Outburst of 2021 Monitored in X-Rays with NICER}",
journal = {ApJ},
year = 2023,
month = sep,
volume = {955},
number = {1},
eid = {37},
pages = {37},
doi = {10.3847/1538-4357/ace9bd},
archivePrefix = {arXiv},
eprint = {2307.11485},
primaryClass = {astro-ph.HE},
adsurl = {https://ui.adsabs.harvard.edu/abs/2023ApJ...955...37O}
}

@ARTICLE{page2022,
author = {{Page}, K.~L. and {Beardmore}, A.~P. and {Osborne}, J.~P. and {Munari}, U. and {Ness}, J. -U. and {Evans}, P.~A. and {Bode}, M.~F. and {Darnley}, M.~J. and {Drake}, J.~J. and {Kuin}, N.~P.~M. and {O'Brien}, T.~J. and {Orio}, M. and {Shore}, S.~N. and {Starrfield}, S. and {Woodward}, C.~E.},
title = "{The 2021 outburst of the recurrent nova RS Ophiuchi observed in X-rays by the Neil Gehrels Swift Observatory: a comparative study}",
journal = {MNRAS},
year = 2022,
month = aug,
volume = {514},
number = {2},
pages = {1557-1574},
doi = {10.1093/mnras/stac1295},
archivePrefix = {arXiv},
eprint = {2205.03232},
primaryClass = {astro-ph.HE},
adsurl = {https://ui.adsabs.harvard.edu/abs/2022MNRAS.514.1557P}
}

@ARTICLE{nelson07,
author = {{Nelson}, T. and {Orio}, M. and {Cassinelli}, J.~P. and {Still}, M. and
{Leibowitz}, E. and {Mucciarelli}, P.},
title = "{X-Ray Spectroscopy of the 2006 Outburst of RS Ophiuchi}",
journal = {ApJ},
eprint = {arXiv:0709.4512},
year = 2008,
month = feb,
volume = 673,
pages = {1067},
doi = {10.1086/524054}
}

@ARTICLE{orio2022,
author = {{Orio}, Marina and {Behar}, E. and {Luna}, G.~J.~M. and {Drake}, J.~J. and {Gallagher}, J. and {Nichols}, J.~S. and {Ness}, J.~U. and {Dobrotka}, A. and {Mikolajewska}, J. and {Della Valle}, M. and {Ignace}, R. and {Rahin}, R.},
title = "{Shocks in the Outflow of the RS Oph 2021 Eruption Observed with X-Ray Gratings}",
journal = {ApJ},
year = 2022,
month = oct,
volume = {938},
number = {1},
eid = {34},
pages = {34},
doi = {10.3847/1538-4357/ac8f4610.48550/arXiv.2209.02123},
archivePrefix = {arXiv},
eprint = {2209.02123},
primaryClass = {astro-ph.HE},
adsurl = {https://ui.adsabs.harvard.edu/abs/2022ApJ...938...34O}
}

@ARTICLE{ness2023,
author = {{Ness}, J.-U. and {Beardmore}, A.~P. and {Bode}, M.~F. and {Darnley}, M.~J. and {Dobrotka}, A. and {Drake}, J.~J. and {Magdolen}, J. and {Munari}, U. and {Osborne}, J.~P. and {Orio}, M. and {Page}, K.~L. and {Starrfield}, S.},
title = "{High-resolution X-ray spectra of RS Ophiuchi (2006 and 2021): Revealing the cause of SSS variability}",
journal = {A\&A},
year = 2023,
month = feb,
volume = {670},
eid = {A131},
pages = {A131},
doi = {10.1051/0004-6361/202245269},
archivePrefix = {arXiv},
eprint = {2212.07169},
primaryClass = {astro-ph.HE},
adsurl = {https://ui.adsabs.harvard.edu/abs/2023A&A...670A.131N}
}

@INPROCEEDINGS{brandi08,
author = {{Brandi}, E. and {Quiroga}, C. and {Ferrer}, O.~E. and {Mikolajewska}, J. and {Garc{\'\i}a}, L.~G.},
title = "{Spectroscopic Orbits and Spectral Variations of RS Ophiuchi}",
booktitle = {RS Ophiuchi (2006) and the Recurrent Nova Phenomenon},
year = 2008,
editor = {{Evans}, A. and {Bode}, M.~F. and {O'Brien}, T.~J. and {Darnley}, M.~J.},
series = {Astronomical Society of the Pacific Conference Series},
volume = {401},
month = dec,
pages = {111},
adsurl = {https://ui.adsabs.harvard.edu/abs/2008ASPC..401..111B}
}

@ARTICLE{bode06,
author = {{Bode}, M.~F. and {O'Brien}, T.~J. and {Osborne}, J.~P. and
{Page}, K.~L. and {Senziani}, F. and {Skinner}, G.~K. and {Starrfield}, S. and
{Ness}, J.-U. and {Drake}, J.~J. and {Schwarz}, G. and {Beardmore}, A.~P. and
{Darnley}, M.~J. and {Eyres}, S.~P.~S. and {Evans}, A. and {Gehrels}, N. and
{Goad}, M.~R. and {Jean}, P. and {Krautter}, J. and {Novara}, G.
},
title = "{Swift Observations of the 2006 Outburst of the Recurrent Nova RS Ophiuchi. I. Early X-Ray Emission from the Shocked Ejecta and Red Giant Wind}",
journal = {ApJ},
eprint = {astro-ph/0604618},
year = 2006,
month = nov,
volume = 652,
pages = {629},
doi = {10.1086/507980}
}

@ARTICLE{rsophshock,
author = {{Ness}, J.-U. and {Drake}, J.~J. and {Starrfield}, S. and {Bode}, M.~F. and {O'Brien}, T.~J. and {Evans}, A. and {Eyres}, S.~P.~S. and {Helton}, L.~A. and {Osborne}, J.~P. and {Page}, K.~L. and {Schneider}, C. and {Woodward}, C.~E.},
title = "{High-Resolution X-Ray Spectroscopy of the Evolving Shock in the 2006 Outburst of RS Ophiuchi}",
journal = {AJ},
year = 2009,
month = feb,
volume = {137},
number = {2},
pages = {3414-3436},
doi = {10.1088/0004-6256/137/2/3414},
archivePrefix = {arXiv},
eprint = {0810.2023},
primaryClass = {astro-ph},
adsurl = {https://ui.adsabs.harvard.edu/abs/2009AJ....137.3414N}
}

@ARTICLE{nessrsoph,
author = {{Ness}, J.-U. and {Starrfield}, S. and {Beardmore}, A.~P. and {Bode}, M.~F. and {Drake}, J.~J. and {Evans}, A. and {Gehrz}, R.~D. and {Goad}, M.~R. and {Gonzalez-Riestra}, R. and {Hauschildt}, P. and {Krautter}, J. and {O'Brien}, T.~J. and {Osborne}, J.~P. and {Page}, K.~L. and {Sch{\"o}nrich}, R.~A. and {Woodward}, C.~E.},
title = "{The SSS Phase of RS Ophiuchi Observed with Chandra and XMM-Newton. I. Data and Preliminary Modeling}",
journal = {ApJ},
year = 2007,
month = aug,
volume = {665},
number = {2},
pages = {1334-1348},
doi = {10.1086/519676},
archivePrefix = {arXiv},
eprint = {0705.1206},
primaryClass = {astro-ph},
adsurl = {https://ui.adsabs.harvard.edu/abs/2007ApJ...665.1334N}
}

@ARTICLE{atel770,
author = {{Osborne}, J. and {Page}, K. and {Goad}, A.~B.~M. and {Bode}, M. and
{O'Brien}, T. and {Schwarz}, G. and {Starrfield}, S. and {Ness}, J.-U. and
{Krautter}, J. and {Drake}, J. and {Evans}, A. and {Eyres}, S.~P.~S.
},
title = "{RS Oph: Swift X-ray observations find short period modulation and highly variable low energy flux}",
journal = {The Astronomer's Telegram},
year = 2006,
month = mar,
volume = 770,
pages = {1}
}

@ARTICLE{basi,
author = {{Ness}, J.~U.},
title = "{High-resolution spectroscopy and high-density monitoring in X-rays of novae}",
journal = {Bulletin of the Astronomical Society of India},
archivePrefix = "arXiv",
eprint = {1209.2153},
primaryClass = "astro-ph.HE",
year = 2012,
month = sep,
volume = 40,
pages = {353},
adsurl = {http://adsabs.harvard.edu/abs/2012BASI...40..353N}
}

@ARTICLE{nessobsc,
author = {{Ness}, J.-U. and {Osborne}, J.~P. and {Henze}, M. and {Dobrotka}, A. and
{Drake}, J.~J. and {Ribeiro}, V.~A.~R.~M. and {Starrfield}, S. and
{Kuulkers}, E. and {Behar}, E. and {Hernanz}, M. and {Schwarz}, G. and
{Page}, K.~L. and {Beardmore}, A.~P. and {Bode}, M.~F.},
title = "{Obscuration effects in super-soft-source X-ray spectra}",
journal = {A\&A},
archivePrefix = "arXiv",
eprint = {1309.2604},
primaryClass = "astro-ph.GA",
year = 2013,
month = nov,
volume = 559,
eid = {A50},
pages = {A50},
doi = {10.1051/0004-6361/201322415},
adsurl = {http://adsabs.harvard.edu/abs/2013A%26A...559A..50N}
}

@ARTICLE{nessqpo,
author = {{Ness}, J.-U. and {Beardmore}, A.~P. and {Osborne}, J.~P. and
{Kuulkers}, E. and {Henze}, M. and {Piro}, A.~L. and {Drake}, J.~J. and
{Dobrotka}, A. and {Schwarz}, G. and {Starrfield}, S. and {Kretschmar}, P. and
{Hirsch}, M. and {Wilms}, J.},
title = "{Short-period X-ray oscillations in super-soft novae and persistent super-soft sources}",
journal = {A\&A},
archivePrefix = "arXiv",
eprint = {1503.00186},
primaryClass = "astro-ph.HE",
year = 2015,
month = jun,
volume = 578,
eid = {A39},
pages = {A39},
doi = {10.1051/0004-6361/201425178},
adsurl = {http://cdsads.u-strasbg.fr/abs/2015A%26A...578A..39N}
}

@ARTICLE{hornebal,
author = {{Horne}, J.~H. and {Baliunas}, S.~L.},
title = "{A prescription for period analysis of unevenly sampled time series}",
journal = {ApJ},
year = 1986,
month = mar,
volume = 302,
pages = {757}
}

@ARTICLE{sssstart_2021,
author = {{Page}, K.~L.},
title = "{RS Oph has entered the Supersoft phase}",
journal = {The Astronomer's Telegram},
year = 2021,
month = sep,
volume = {14894},
pages = {1},
adsurl = {https://ui.adsabs.harvard.edu/abs/2021ATel14894....1P}
}

@ARTICLE{swnovae,
author = {{Ness}, J.U. and {Schwarz}, G.~J. and {Retter}, A. and {Starrfield}, S. and
{Schmitt}, J.~H.~M.~M. and {Gehrels}, N. and {Burrows}, D. and
{Osborne}, J.~P.},
title = "{Swift X-Ray Observations of Classical Novae}",
journal = {ApJ},
eprint = {arXiv:astro-ph/0703286},
year = 2007,
month = jul,
volume = 663,
pages = {505},
doi = {10.1086/518084}
}

@ARTICLE{mitrani2024,
author = {{Mitrani}, Sharon and {Behar}, Ehud and {Drake}, Jeremy J. and {Orio}, Marina and {Page}, Kim and {Canton}, Valentina and {Ness}, Jan-Uwe and {Sokolovsky}, Kirill},
title = "{Mixing of Hot Shocked Plasma with Cold Gas in Nova YZ Ret 2020}",
journal = {ApJ},
year = 2024,
month = jul,
volume = {970},
number = {1},
eid = {54},
pages = {54},
doi = {10.3847/1538-4357/ad4a64},
archivePrefix = {arXiv},
eprint = {2405.09219},
primaryClass = {astro-ph.HE},
adsurl = {https://ui.adsabs.harvard.edu/abs/2024ApJ...970...54M}
}

@ARTICLE{mitrani_v1716sco,
       author = {{Mitrani}, Sharon and {Behar}, Ehud and {Orio}, Marina and {Worley}, Jack},
        title = "{X-Ray Observations of Nova Sco 2023: Spectroscopic Evidence of Charge Exchange}",
      journal = {ApJ},
         year = 2025,
        month = aug,
       volume = {989},
       number = {2},
          eid = {166},
        pages = {166},
          doi = {10.3847/1538-4357/adf1a3},
archivePrefix = {arXiv},
       eprint = {2507.02465},
 primaryClass = {astro-ph.HE},
       adsurl = {https://ui.adsabs.harvard.edu/abs/2025ApJ...989..166M}
}

@ARTICLE{rohrbach09,
       author = {{Rohrbach}, J.~G. and {Ness}, J.-U. and {Starrfield}, S.},
        title = "{Evolution of X-Ray Spectra and Light Curves of V1494 Aquilae}",
      journal = {AJ},
         year = 2009,
        month = jun,
       volume = {137},
       number = {6},
        pages = {4627-4632},
          doi = {10.1088/0004-6256/137/6/4627},
archivePrefix = {arXiv},
       eprint = {0903.5386},
 primaryClass = {astro-ph.HE},
       adsurl = {https://ui.adsabs.harvard.edu/abs/2009AJ....137.4627R}
}
